\documentclass[12pt]{iopart}

\usepackage{iopams}
\usepackage{graphicx}

\begin{document}

\title{Vortex Knots in Light}

\author{J Leach\dag\footnote[3]{To whom correspondence should be addressed.}, M R Dennis\ddag, J Courtial\dag, and M J Padgett\dag}

\address{\dag Physics and Astronomy, University Av, Kelvin Bld, Glasgow G12 8QQ, UK}

\address{\ddag School of Mathematics, University of Southampton, Highfield SO17 1BJ, UK}

\begin{abstract}

Optical vortices generically arise when optical beams are combined. 
Recently, we reported how several laser beams containing optical vortices could be combined to form optical vortex loops, links and knots embedded in a light beam (Leach \etal 2004 {\em Nature} {\bf 432} 165).  
Here, we describe in detail the experiments in which vortex loops form these structures.  
The experimental construction follows a theoretical model originally proposed by Berry and Dennis, and the beams are synthesised using a programmable spatial light modulator and imaged using a CCD camera.

\end{abstract}

\pacs{03.65.Vf, 02.10.Kn, 42.25.Hz, 42.40.Jv} 


Optical vortices (phase singularities, wave dislocations) are ubiquitous in nature, occurring whenever three or more light waves interfere.  
They are places in the interference field where the intensity is zero, and the phase is undefined, and usually occur at points in 2-dimensional fields, and along lines in 3 dimensions (see, for example, \cite{nb:34, bnw:79, nye:natural}).  
Within the laboratory, vortices are commonly produced using computer generated holograms or diffractive optics.  
Holograms of the design originally demonstrated by \cite{Bazhenov, hmsw:generation} are now widely used to create light beams with vortices embedded, and the orbital angular momentum \cite{absw:laguerre, apb:oam, abp:oam} with which they are associated.  
In 2001, Berry and Dennis showed theoretically \cite{bd:332, bd:333} that specific superpositions of beams could be generated in which vortex loops could be linked together and even knotted. 
Here, we describe in detail our experimental generation \cite{ldcp:knotted} of such laser beams.

An optical vortex embedded on the axis of a laser beam is usually characterised by a function, $R^{|l|}\exp(\rmi l \phi)$ describing the local structure of the optical field around the vortex point in a transverse cross section of the beam. 
The $\exp(\rmi l \phi)$ factor describes the phase singularity nature of the vortex, $R^{|l|}$ shows that this must be accompanied by a zero of intensity. 
The integer $l$ determines the number of phase cycles made in a right-handed circuit of the vortex and is called the strength of the singularity.  
Although $l$ can take any integer value, positive or negative, vortices with $|l|>1$ are unstable and, in the presence of astigmatism, degenerate to form multiple vortices of $l = \pm 1$ upon propagation.  
Many observations have been reported on the seeming interaction between two or more vortex points as light propagates from one plane to the next (see, for instance, \cite{rls:propagation} and references therein). 
In this view, two vortices with oppositely signed $l$ may spontaneously appear together in one plane, propagate to a subsequent plane, and then recombine (birth/creation and death/annihilation of vortices).  
By identifying propagation of the beam with temporal evolution, this language is possibly misleading since the optical beam is static in time in the laboratory reference frame. 

A different physical description is to recognise that the pair of vortices forms a looped thread within the beam (see, for example, \cite{berry:296}).  
An arbitrary light field, such as a space-filling speckle pattern, contains many optical vortex lines in a complicated arrangement \cite{bd:321}. 
As parameters vary, the vortex lines evolve in a complicated way, and may reconnect, and loops may appear or disappear \cite{nb:34, bd:333, nye:local, dennis:thesis}.

The original analysis of knotted and linked optical vortex loops \cite{bd:332} used a combination of Bessel beams \cite{dme:diffraction}.
From an experimental standpoint this construction has two problems: firstly, that ideal Bessel beams have infinite aperture (their transverse profile, specified by a $J_{l}$ Bessel function, is infinitely extended); secondly, that the range of intensity in the proposed beams was too large to observe the faint light around the knotted and linked dark threads.  
The construction was later extended \cite{bd:333} to give the same topological structures within paraxial beams, allowing the possibility of knotted dark structures within Laguerre-Gaussian beams \cite{absw:laguerre}, with fixed frequency specified by the wavenumber $k.$ 
In cylindrical coordinates $R, \phi, z,$ these have the normalised form
\begin{eqnarray}
\psi_{lp}(R, \phi, z; w) & = & \sqrt{\frac{p!}{\pi(|l|+p)!}}\frac{R^{|l|}\exp(\rmi l \phi)}{(w^2+\rmi z / k)^{|l|+1}} \exp\left(\frac{-R^2}{2(w^2 + \rmi z/k)}\right) \nonumber\\
 &  & \times L_p^{|l|}\left(\frac{R^2}{w^2+z^2/k^2 w^2}\right)\left(\frac{w^2 - \rmi z / k}{w^2 + \rmi z / k}\right)^p \label{eq:lg}
\end{eqnarray}
where $L$ denotes the generalized Laguerre polynomials. 
In addition to an axial optical vortex of strength $l$, there are two other parameters that may be varied - the radial number $p,$ which is an integer determining the order of the Laguerre polynomial (effectively, the number of nodal rings about the beam axis), and the waist width $w,$ which is a continuous variable setting the finite transverse width of the beam.  
These additional degrees of freedom allow us to create the same topological structures, but with a reduced range of intensity, enabling the nodal structures to be unambiguously seen.

We synthesize the link and knot by superposing three Laguerre-Gaussian beams with the same index $l,$ giving the field $\psi_{\mathrm{unpert}},$ then perturbed by another Gaussian beam  $\psi_{\mathrm{pert}}$ with index $l = 0.$ 
Therefore, the knot and link fields $\psi_{\mathrm{knot,link}}$ are equal to the sum $\psi_{\mathrm{unpert}}+\psi_{\mathrm{pert}}.$ 
The amplitudes, values of $p$ and waist width $w$ of the components of the unperturbed superposition $\psi_{\mathrm{unpert}}$ are chosen so that at a specific radius $R_0$ in the waist plane, both the field and its radial derivative are zero, i.e.
\begin{equation}
\psi_{\mathrm{unpert}}(R_{0}) = 0, \qquad \partial_{R} \psi_{\mathrm{unpert}}|_{R_{0}} = 0. \label{eq:conds}
\end{equation}

The resulting destructive interference forms a configuration of three circular vortex loops in the transverse plane, centred on the $z$-axis.  
One is in the waist plane ($z=0$) with radius $R_{0},$ and the other two, which share the same smaller radius, are symmetrically arranged on either side of this plane.  
The fact that the radial derivative vanishes on the central vortex ring ($z = 0, R = R_0$) means that it has zero strength - it can be thought of as two superposed rings with opposite strengths.  
The numerical value of $R_0$ and particular choices of $p$ and $w$ in the superposed beams are chosen such that the intensity is relatively bright in the regions separating the three vortex rings from each other and from the $l$-fold singularity on the axis.

\begin{figure}
\centerline{\includegraphics{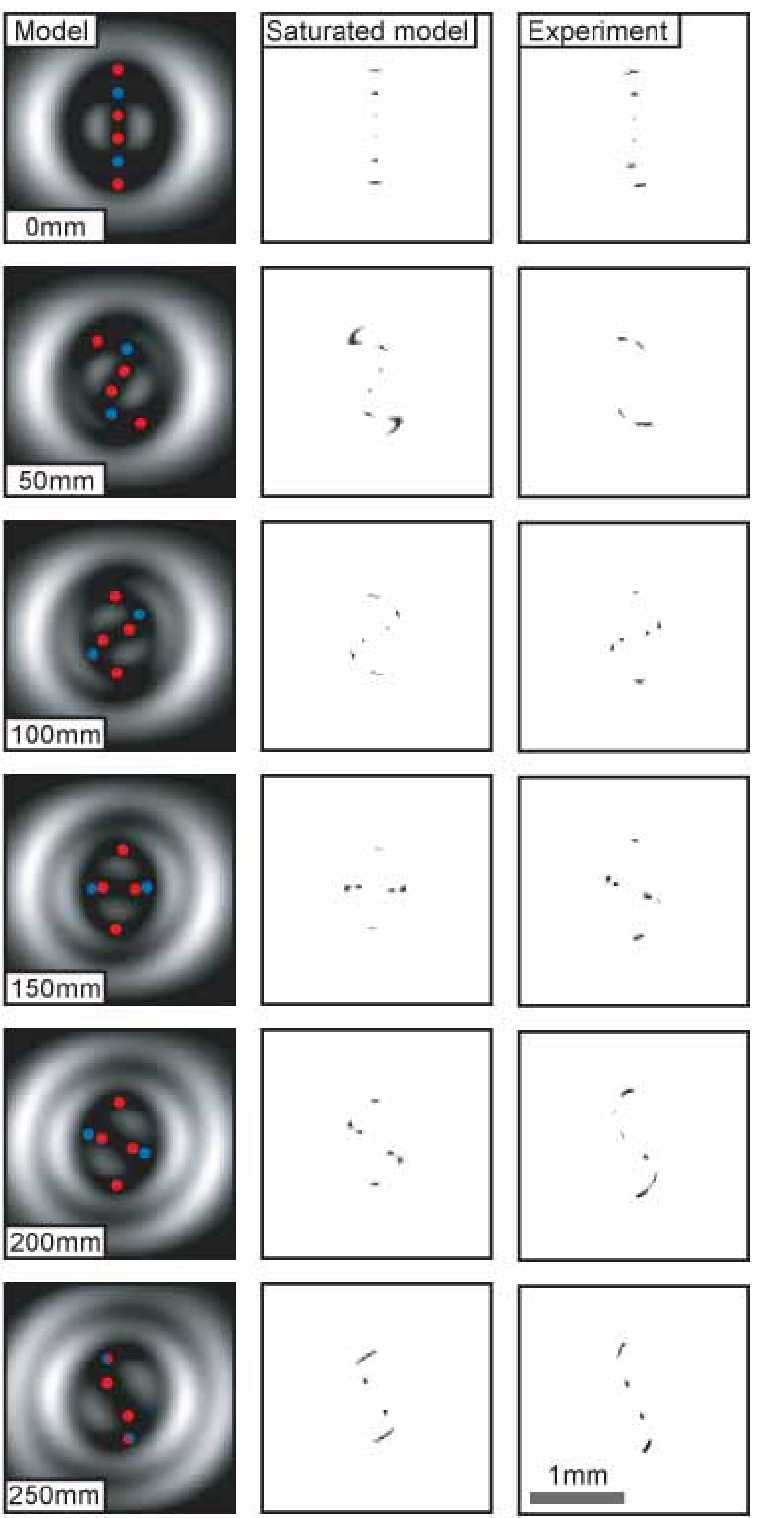}} 
\caption{
Theoretical and experimental transverse cross-sections of the vortex link.  
The distances refer to the distance from the beam waist.  
In the model, the positions of the vortices are indicated by coloured spots, red $(l = + 1)$ and blue $(l = - 1)$. 
These vortices are observed as black dots in the saturated model and the experimental results.} 
\label{fig:Link2d}
\end{figure}

\begin{figure}
\centerline{\includegraphics{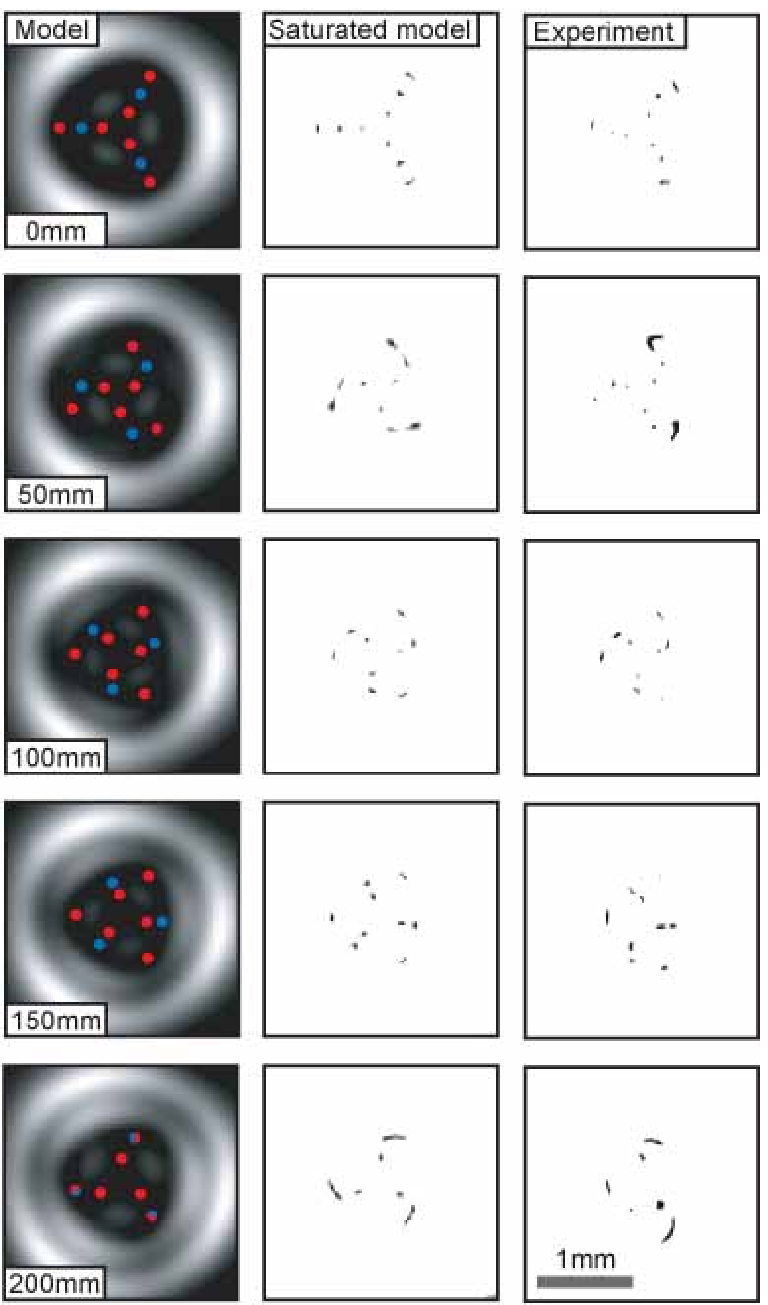}} 
\caption{
Theoretical and experimental transverse cross-sections of the vortex knot.  The distances refer to the distance from the beam waist.  In the model, the positions of the vortices are indicated by coloured spots, red $(l = + 1)$ and blue $(l = - 1)$.  These vortices are observed as black dots in the saturated model and the experimental results.}
\label{fig:Knot2d}
\end{figure}

\begin{figure}
\centerline{\includegraphics{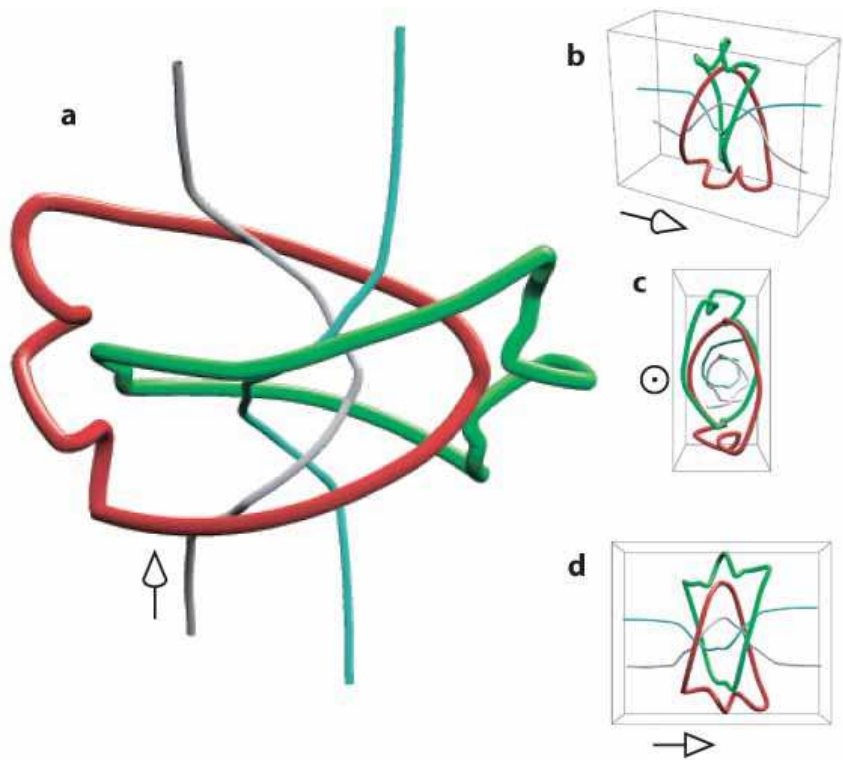}} 
\caption{
3-dimensional representation of the experimental vortex link, reconstructed from the measured data.  
Different projections of the link are shown, and the direction of propagation in each case is indicated by the arrow.  
The link is formed by the red and green loops.  
The threading vortices, which do not take part in the link, are represented by thinner blue and grey lines.  
The scale for the direction of propagation has been reduced by approximately 1000. } 
\label{fig:Link3d}
\end{figure}

\begin{figure}
\centerline{\includegraphics{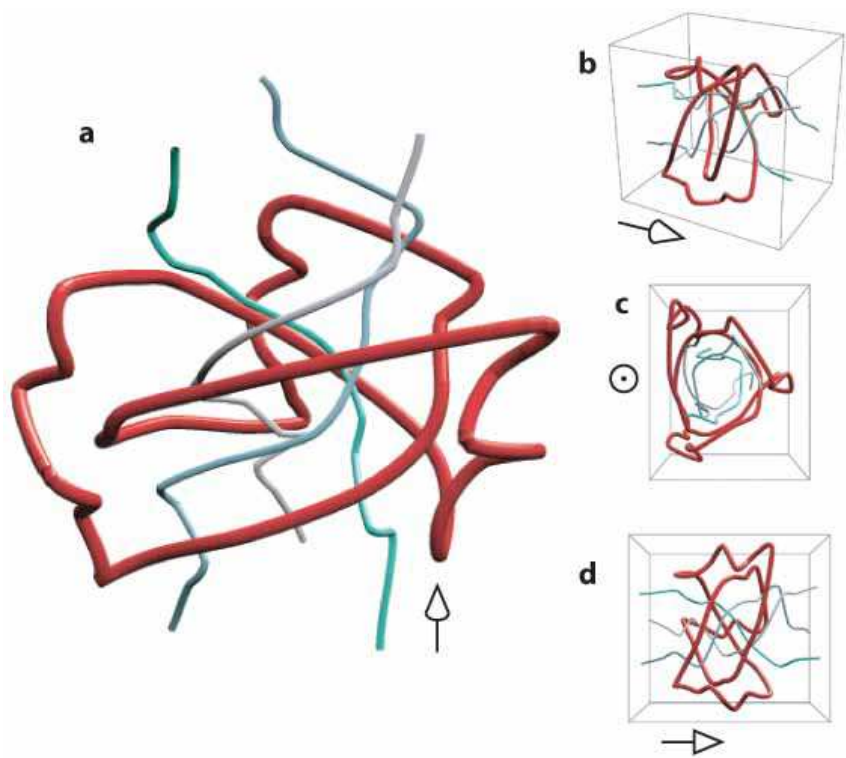}} 
\caption{
3-dimensional representation of the experimental vortex trefoil knot, reconstructed from the measured data.  
Different projection of the knot are shown, and the direction of propagation in each case is indicated by the arrow.  
The knot is represented by the red loop.  
The threading vortices, which do not take part in the knot, are represented by thinner blue and grey lines. 
The scale for the direction of propagation has been reduced by approximately 1000. } 
\label{fig:Knot3d}
\end{figure}

To this superposition is added a simple Gaussian perturbing beam $\psi_{\mathrm{pert}}$ with near planar phasefronts (i.e. wide waist).  
As the amplitude of the perturbing beam increases, the configuration of three vortex rings deforms through a sequence of reconnections (explained in detail in \cite{bd:333}) to give an $(l, 2)$ torus knot \cite{adams:knot}.  
Therefore, the total superposition of beams we used is 
\begin{eqnarray}
   \psi_{\mathrm{link,knot}}(x,y,z) & = & \psi_{\mathrm{unpert}} + \psi_{\mathrm{pert}} \nonumber \\
   & = & a_1 \psi_{l0}(x,y,z;2w_0) + a_2 \psi_{l0}(x,y,z;w_0) \nonumber \\
   & &  + a_3 \psi_{l1}(x,y,z;w_0) + a_{\mathrm{pert}} \psi_{00}(x,y,z;8w_0)
   \label{eq:superposn}
\end{eqnarray}
The superposition of beams with amplitudes $a_{1,2,3}$ form $\psi_{\mathrm{unpert}},$ satisfying the conditions (\ref{eq:conds}) above.
For convenience, $a_1 = 1,$ and $w_0$ is a fixed transverse lengthscale of the beam, chosen in our experiment to be 1mm. 
The link, depicted later in figures \ref{fig:Link2d} and \ref{fig:Link3d}, was formed with $l = 2,$ and the relative amplitudes: $a_2 = -0.35, a_3 = 0.36, a_{\mathrm{pert}} = 0.35$ (which gave $R_0 = 2w_0$).  
The knot of figures \ref{fig:Knot2d} and \ref{fig:Knot3d} was formed with $l = 3,$ and relative amplitudes $a_2 = -0.26, a_3 = 0.29, a_{\mathrm{pert}} = 0.25$ (which gave $R_0 = 2.2 w_0$).  
If $l = 1$ the `torus knot' is a single loop, wrapped twice around the vortex axis, which will not be described further.  
The beam combinations we have chosen, satisfying the general conditions of \cite{bd:333}, are not unique (two equations are solved, but three parameters ($a_2, a_3$ and $R_0$) are varied; also, the values of $p$ and $w$ can be changed).  
They are convenient in that they are straightforward to implement and the resulting intensity range makes it possible to image the singularities. 
Adjustment of the superposition numerically reveals that that the link occurs for values of the relative perturbation amplitude $a_{\mathrm{pert}}$ between 0.3 and 0.42.  
Both the knot and link are stable between similar finite ranges of their various parameters.

As discussed above, vortex carrying beams can be easily created within the laboratory using diffractive optical components.  
For example, if a diffraction grating is modified such that the lines near its centre combine to form a fork dislocation, then the first-order diffracted beam contains an optical vortex.  
The strength $l$ of the vortex is equal to the order of the fork dislocation.  
In addition, $p$ circular discontinuity or discontinuities can be added to these designs to give the radial index of the resulting beam \cite{arlt:98}.  
Although a hologram of this kind provides precise control over the phase of the diffracted beam, the intensity distribution depends also on the profile of the illumination beam.  
It transpires that for a simple single-forked hologram design, illumination by a fundamental Gaussian mode gives an $l = 1$ vortex beam in the far field which has over 80\% of its energy in the $l = 1, p = 0$ Laguerre-Gaussian mode.  
Although satisfactory for many studies and applications, this level of mode purity is insufficient to form the topological structure we desire.

The design of phase-only holograms can be modified to control not only the phase structure of the diffracted beams but their intensity also \cite{vasara:89, basistiy:03, franke-arnold:04}. 
In essence the intensity at any point in a plane can be attenuated by adjusting the efficiency of the blazing in that region of the hologram.

Our required superposition of Laguerre-Gaussian beams could be generated with multiple holograms, but it is more convenient to produce a single diffractive component that produces the combination of beams directly.  
We adopt a simple yet highly effective approach to phase-only hologram design.  
Once the required indices, amplitude weightings and relative phases of the ingredient modes are determined theoretically (as described above), the phase distribution of the link and knot beam in the waist plane, $\Phi (x, y)_{\mathrm{beam}} = \arg \psi_{\mathrm{link,knot}}(x,y,0),$ is calculated.  
To this phase distribution is added the phase distribution of a blazed diffraction grating $\Phi (x, \Lambda)_{\mathrm{grating}}$ (where $\Lambda$ is the period of the grating), such that the first order diffracted energy is angularly separated from the other orders, which are subsequently blocked using a spatial filter.  
The desired intensity $I(x, y)$ of the superposition is also calculated and normalised so the maximum intensity is unity.  
This intensity distribution is applied as a multiplicative mask to the phase distribution of the hologram, acting as a selective beam attenuator imposing the necessary intensity distribution on the first order diffracted beam.

\begin{figure}
\centerline{\includegraphics{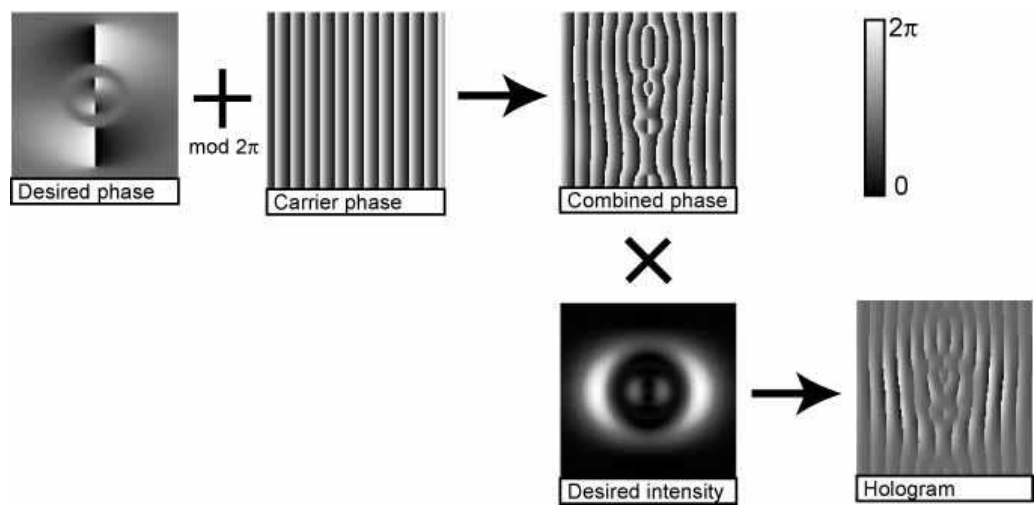}} 
\caption{
Illustration of the algorithm used to generate the holograms used in the link experiment (the knot hologram was similar).  
The desired phase of the beam at the beam waist is added mod$2\pi$ to a carrier. 
This is then multiplied by the desired intensity of the beam (corrected to account for the mapping of phase depth to diffraction efficiency) to produce the hologram which is used in the experiment. 
The 0-$2\pi$ range is represented in greyscale, as is actually programmed into the hologram.} 
\label{fig:Hologram}
\end{figure}

The resulting phase distribution of the hologram $\Phi (x, y)_{\mathrm{holo}}$, expressed within a 0 to $2 \pi$ range is given by
\begin{eqnarray}
\Phi (x, y)_{\mathrm{holo}} & = &
 \left( \left( \left( \Phi (x, y)_{\mathrm{beam}} + \Phi (x, \Lambda)_{\mathrm{grating}} \right)_{\mathrm{mod} 2 \pi} - \pi \right) \right) \nonumber \\
 &  &  \times \mathrm{sinc}^2 \left( \left(1 - I(x,y) \right) \pi \right) + \pi. \label{eq:holo}
\end{eqnarray}
The sinc distortion of the intensity terms accounts for the mapping of phase depth to diffraction efficiency.  
An example of this process is shown in figure \ref{fig:Hologram}.

To implement this hologram design we use a programmable spatial light modulator (HoloEye LC-R3000) under computer control to produce the desired beam.
Such devices are based on liquid crystal arrays, which introduce spatially dependent phase retardation across the beam profile, allowing the creation of any specific combination of beams.  
Compared to a static hologram, the spatial light modulator allows us to control the beam parameters dynamically and, most importantly, the amplitude of the perturbation in real time.

\begin{figure}
\centerline{\includegraphics{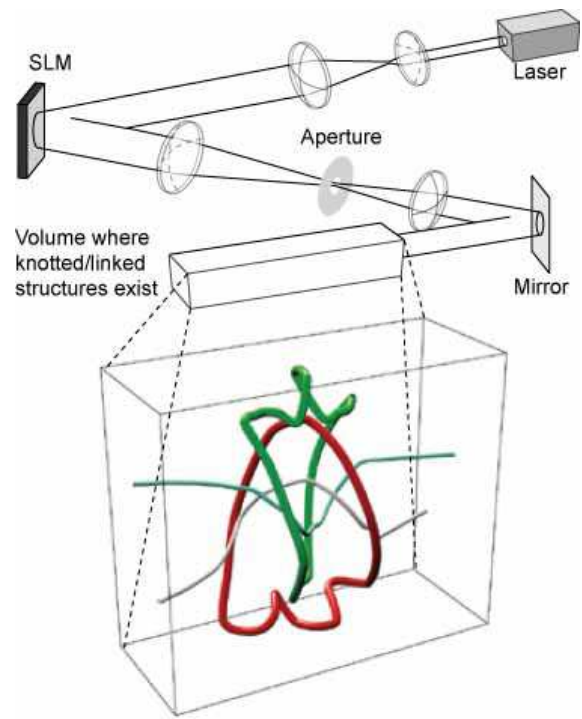}} 
\caption{
Experimental configuration used to produce the vortex beams.  
The expanded and collimated beam from a He-Ne laser is reflected by the spatial light modulator displaying the required hologram.  
The beam is passed through a lens to spatially separate the diffraction orders. 
The first diffracted order is selected, and the other orders removed, by passing the light through an adjustable iris.  
A second lens is used to recollimate the beam and allow transverse cross-sections of the beam to be imaged.  
A CCD camera is placed at different planes to record the intensity of the beam.} 
\label{fig:Experiment}
\end{figure}

The optical beams were produced using a spatial filtered He-Ne laser, expanded to overfill the 20mm aperture of the spatial light modulator, and configured to generate the appropriate superposition of Laguerre-Gaussian beams, as shown in figure \ref{fig:Experiment}.

\begin{figure}
\centerline{\includegraphics{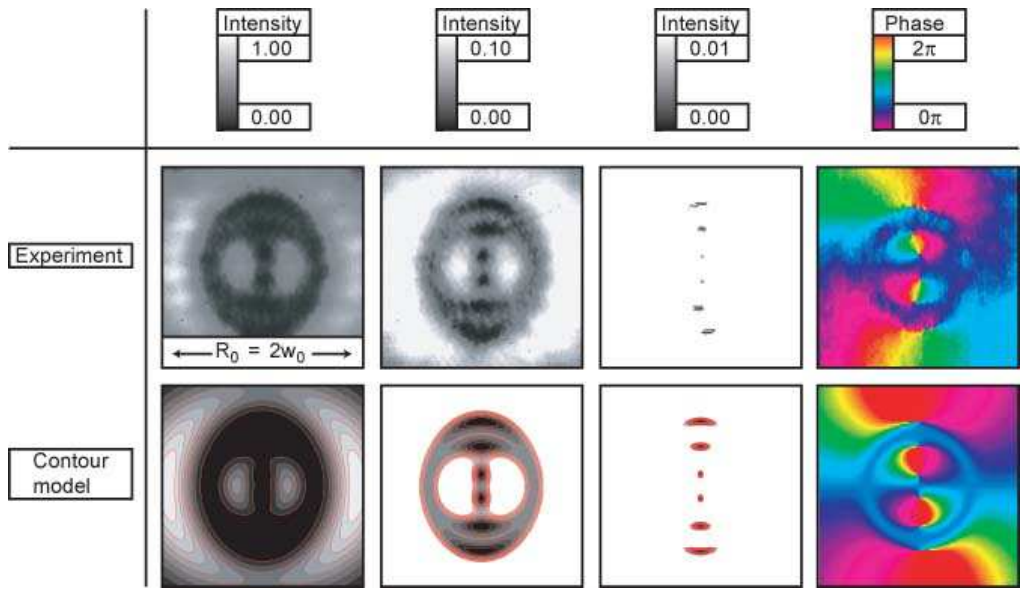}} 
\caption{
Experimental and theoretical representation of the transverse structure of the waist of the link field.
A sequence of plots of intensity, with increasing degree of saturation, shows how the only structure measured are the vortex lines piercing the observation plane.
The phase (conventionally represented using hue), shows the cyclic phase change around the vortices, establishing the singular phase nature of the intensity nodes.} 
\label{fig:Saturation}
\end{figure}

The hologram was calculated at 800 by 800 pixels and displayed on the SLM at 1200 by 1200 pixels.  
The period of the diffraction grating was 0.4 lines/mm such that, after passing through a lens, the first order diffracted energy was angularly separated from the other orders by approximately 2mm at the Fourier plane of the hologram.  
The unwanted diffraction orders were subsequently blocked by centering the first diffracted order on an adjustable iris which acted as a spatial filter.  
The size of the spatial filter was such that diffraction effects were negligible.  
A second lens was used to create an image of the modulator.  
A CCD array positioned in this second plane and translated along the beam axis measured the intensity structure of the beam on each side of the waist.  
Given that the fine details of the vortex threads lie in regions of near darkness and in order to measure the vortex positions within the beam cross-section, it was necessary to over-saturate the CCD.
The resulting image was mainly white, but contains points of darkness corresponding to the vortices as they intersected the plane of the CCD.  
A comparison of experimental and theoretical plots demonstrating the increasing saturation of the detector in the waist plane of the link field are shown in figure \ref{fig:Saturation}.
Also shown here is the measured phase pattern of the beam, showing that the dark vortex threads are indeed singularities of the optical phase (this phase distribution was measured according to the methods described in \cite{lyp:observation}.

These transverse vortex coordinates were measured in neighbouring planes, enabling the 3-dimensional spatial configuration to be deduced.  
The amplitude of the perturbation was chosen initially at the theoretically optimized value, then adjusted to give the best defined link or knot.  
In figures \ref{fig:Link2d} and \ref{fig:Knot2d}, images taken at various transverse planes through the vortex link and knot beams are shown alongside their simulated counterparts.
The measured positions of the vortices are extremely close to those determined theoretically for the modelled beams.

Figures \ref{fig:Link3d} and \ref{fig:Knot3d} represent the experimental spatial vortex link and knot constructions reconstructed from the measurements. 
Represented in this way, it is clear that the vortex topologies are as claimed, although the vortex loops making them are not particularly smooth. 
This is a consequence of the choice of superposed beams, here chosen so that the range of intensity in the vicinity of the vortex loops is small enough to pick out the vortices.
In other constructions, such as that in \cite{bd:332} using Bessel beams, the knot and link curves are smoother, but sit in regions of near-darkness.
It is possible that smooth vortex knots may exist in constructions similar to ours with a reduced range of intensity, but we were unable to find these.

We have described the creation and observation of static knots and links of dark vortex threads in light waves, verifying an earlier prediction of Berry and Dennis that this was possible within optical wave fields.  
Knotted vortices have been studied theoretically since Lord Kelvin's vortex atom hypothesis \cite{kelvin:atoms} in a range of different physical situations, such as hydrodynamics \cite{moffatt:knottedness, az:linking}, field theory \cite{fn:stable} and nonlinear excitable media \cite{ws:filaments3, sw:stability}. 
However, in these contexts, nonlinearities mean that vortex knots are probably unstable.  
Our structures are superpositions of free space optical modes with the same frequency, so are temporally stable.  
It is not known theoretically whether arbitrary topological configurations of vortices can be embedded in free space optical fields; only a few special cases are known (the knots and links studied here \cite{bd:332, bd:333}, which have to be threaded, and certain braids \cite{dennis:braided}, which require counterpropagating beams).  
In addition to being of fundamental interest, these experiments illustrate topological light shaping that might be used as a topological waveguide for quantum mechanical matter waves, such as Bose-Einstein condensates \cite{ra:creating,dr:transfer}.

\ack

We are grateful to Michael Berry for useful suggestions and comments. 
Figures \ref{fig:Link3d} and \ref{fig:Knot3d} were plotted in Mathematica using Mitch Berger's {\tt tuba} package.  
MRD acknowledges support from the Leverhulme Trust and the Royal Society.
JC is supported by the Royal Society.

\section*{References}


\begin{thebibliography}{25}


\bibitem{nb:34}
Nye J F and Berry M V 1974
\newblock Dislocations in wave trains
\newblock {\em Proc. R. Soc. Lond. A} {\bf 336} 165--90

\bibitem{bnw:79}
Berry M V, Nye J F and Wright F J 1979
\newblock The elliptic umbilic diffraction catastrophe
\newblock {\em Phil. Trans. R. Soc. A} {\bf 291} 453--84

\bibitem{nye:natural}
Nye J F 1999
\newblock {\em Natural focusing and fine structure of light: caustics and wave dislocations}.
\newblock Bristol: Institute of Physics Publishing

\bibitem{Bazhenov}
Bazhenov V Y, Vasnetsov M V and Soskin M S 1990
\newblock Laser beams with screw dislocations in their wavefronts
\newblock  {\em JEPT Lett.}  {\bf 52} 429Ð-31

\bibitem{hmsw:generation}
Heckenberg N R, McDuff R, Smith C P and White A G 1992
\newblock Generation of optical-phase singularities by computer-generated holograms 
\newblock {\em Opt. Lett.} {\bf 17} 221--23

\bibitem{absw:laguerre}
Allen L, Beijersbergen M, Spreeuw R J C, and Woerdman J P 1992
\newblock Orbital angular momentum of light and the transformation of Laguerre-Gaussian laser modes
\newblock \PR {\em A} {\bf 45} 8185--9

\bibitem{apb:oam}
Allen L, Padgett M J, and Babiker M 1999
\newblock The orbital angular momentum of light
\newblock {\em Prog. Opt.} {\bf 39} 291--372

\bibitem{abp:oam}
Allen L, Barnett S M, and Padgett M J eds 2003
\newblock The orbital angular momentum of light
\newblock Bristol: Institute of Physics Publishing

\bibitem{bd:332}
Berry M V and Dennis M R 2001
\newblock Knotted and linked phase singularities in monochromatic waves
\newblock {\em Proc. R. Soc. Lond. A} {\bf 457} 2251--63

\bibitem{bd:333}
Berry M V and Dennis M R 2001
\newblock Knotting and unknotting of phase singularities: Helmholtz waves, paraxial waves and waves in 2+1 dimensions
\newblock \JPA {\bf 34} 8877--88

\bibitem{ldcp:knotted}
Leach J, Dennis M R, Courtial J, and Padgett M J 2004
\newblock Knotted threads of darkness
\newblock {\em Nature} {\bf 432} 165

\bibitem{rls:propagation}
Rozas D, Law C T, and Swartzlander G A 1997
\newblock Propagation dynamics of optical vortices
\newblock {\em J. Opt. Soc. Am. B} {\bf 14} 3054--65

\bibitem{berry:296}
Berry M V 1998
\newblock Much ado about nothing: optical dislocation lines (phase singularities, zeros, vortices...),
\newblock in {\em Proceedings of International Conference on Singular Optics}, Soskin M S ed, SPIE volume {\bf 3487} 1--15

\bibitem{bd:321}
Berry M V and Dennis M R 2000
\newblock Phase singularities in isotropic random waves
\newblock {\em Proc. R. Soc. Lond. A} {\bf 456} 2059--79

\bibitem{nye:local}
Nye J F 2004
\newblock Local solutions for the interaction of wave dislocations
\newblock \JOA {\bf 6} S251-4

\bibitem{dennis:thesis}
Dennis M R 2001
\newblock {\em Topological singularities in wave fields}
\newblock PhD thesis, Bristol University

\bibitem{dme:diffraction}
Durnin J, Miceli~Jr J J, and Eberly J H 1997
\newblock Diffraction-free beams
\newblock \PRL {\bf 58} 1499--501

\bibitem{adams:knot}
Adams C C 1994
\newblock {\em The knot book}
\newblock New York: Freeman

\bibitem{arlt:98}
Arlt J, Dholakia K, Allen L, and Padgett M J 1998
\newblock The production of multiringed Laguerre-Gaussian modes by computer-generated holograms
\newblock {\em J. Mod. Opt.} {\bf 45} 1231--7

\bibitem{vasara:89}
Vasara A, Turunen J, and Friberg A T 1989
\newblock Realization of general nondiffracting beams with computer-generated holograms
\newblock {\em J. Opt. Soc. Am. A} {\bf 6} 1748--54

\bibitem{basistiy:03}
Basistiy I V, Slyusar V V, Soskin M S, Vasnetsov M V, and Bekshaev A Y 2003
\newblock Manifestation of the rotational Doppler effect by use of an off-axis optical vortex beam
\newblock {\em Opt. Lett.} {\bf 28} 1185-7

\bibitem{franke-arnold:04}
Franke-Arnold S, Barnett S M, Yao E, Leach J, Courtial J and Padgett M J 2004
\newblock Uncertainty principle for angular position and angular momentum
\newblock \NJP {\bf 6} 103
 
\bibitem{lyp:observation}
Leach J, Yao E and Padgett M J 2004
\newblock Observation of the vortex structure of a non-integer vortex beam
\newblock \NJP {\bf 6} 71

\bibitem{kelvin:atoms}
Lord Kelvin (W Thompson) 1867
\newblock On vortex atoms
\newblock {\em Phil. Mag.} {\bf 34} 15--24

\bibitem{moffatt:knottedness}
Moffatt H K 1969
\newblock The degree of knottedness of tangled vortex lines
\newblock {\em J. of Fluid Mech.} {\bf 35} 117--29

\bibitem{az:linking}
Aref H and Zawadzki I 1991
\newblock Linking of vortex rings
\newblock {\em Nature} {\bf 354} 50--3

\bibitem{fn:stable}
Faddeev L and Niemi A J 1997
\newblock Stable knot-like structures in classical field theory
\newblock {\em Nature} {\bf 387} 58--61

\bibitem{ws:filaments3}
Winfree A T and Strogatz S H 1983
\newblock Singular filaments organize chemical waves in three dimensions. III.
  Knotted waves
\newblock {\em Physica D} {\bf 9} 333--45

\bibitem{sw:stability}
Sutcliffe P M and Winfree A T 2003
\newblock Stability of knots in excitable media
\newblock \PR {\em E} {\bf 68} 016218

\bibitem{dennis:braided}
Dennis M R 2003
\newblock Braided nodal lines in wave superpositions
\newblock \NJP {\bf 5} 134

\bibitem{ra:creating}
Ruostekoski J and Anglin J R 2001
\newblock Creating vortex rings and three-dimensional skyrmions in Bose-Einstein condensates
\newblock \PRL {\bf 86} 3934--7

\bibitem{dr:transfer}
Dutton Z and Ruostekoski J 2004
\newblock Transfer and storage of vortex states in light and matter waves
\newblock \PRL {\bf 93} 193602


\end{thebibliography}

\newpage

\newpage

\newpage

\newpage

\newpage

\newpage

\newpage

\end{document}